\documentclass[%
 reprint,
superscriptaddress,
bibnotes,
amsmath,amssymb,
]{revtex4-2}

\usepackage[utf8]{inputenc}
\usepackage{siunitx}
\usepackage[english]{babel}
\usepackage{graphicx}
\usepackage{amssymb,amsmath}
\usepackage{float}
\usepackage{braket}
\usepackage{bm}
\usepackage{hyperref}
\hypersetup{colorlinks=true,allcolors=blue}
\usepackage{lineno}

\begin{document}

\date{\today}
\title{Experimental anonymous quantum conferencing}

\author{Jonathan W. Webb}
\affiliation{Institute of Photonics and Quantum Sciences, School of Engineering and Physical Sciences, Heriot-Watt University, Edinburgh, EH14 4AS, UK}
\author{Joseph Ho}
\affiliation{Institute of Photonics and Quantum Sciences, School of Engineering and Physical Sciences, Heriot-Watt University, Edinburgh, EH14 4AS, UK}
\author{Federico Grasselli}
\affiliation{Institut für Theoretische Physik III, Heinrich-Heine-Universität Düsseldorf, Universitätsstraße 1, D-40225 Düsseldorf, Germany}
\author{Gl\'aucia Murta}
\affiliation{Institut für Theoretische Physik III, Heinrich-Heine-Universität Düsseldorf, Universitätsstraße 1, D-40225 Düsseldorf, Germany}
\author{Alexander Pickston}
\affiliation{Institute of Photonics and Quantum Sciences, School of Engineering and Physical Sciences, Heriot-Watt University, Edinburgh, EH14 4AS, UK}
\author{Andr\'es Ulibarrena}
\affiliation{Institute of Photonics and Quantum Sciences, School of Engineering and Physical Sciences, Heriot-Watt University, Edinburgh, EH14 4AS, UK}
\author{Alessandro Fedrizzi}
\email{a.fedrizzi@hw.ac.uk}
\affiliation{Institute of Photonics and Quantum Sciences, School of Engineering and Physical Sciences, Heriot-Watt University, Edinburgh, EH14 4AS, UK}
\begin{abstract}
Anonymous quantum conference key agreement (AQCKA) allows a group of users within a network to establish a shared cryptographic key without revealing their participation.
Although this can be achieved using bi-partite primitives alone, it is costly in the number of network rounds required. By allowing the use of multi-partite entanglement, there is a substantial efficiency improvement.
We experimentally implement the AQCKA task in a six-user quantum network using Greenberger-Horne-Zeilinger (GHZ)-state entanglement and obtain a significant resource cost reduction in line with theory when compared to a bi-partite-only approach.
We also demonstrate that the protocol retains an advantage in a four-user scenario with finite key effects taken into account.
\end{abstract}

\maketitle

\section*{introduction}

A wide range of modern communication tasks involve group settings where multiple parties communicate, e.g. in a conference call, or in a sensor network~\cite{ingemarsson_conference_1982,blundo_perfectly-secure_1993, burmester_secure_1995, guo_distributed_2020,komar_quantum_2014-1,liu_distributed_2021-1,tubaishat_sensor_2003}.
These group sessions can be secured cryptographically, in addition one may require anonymity where sub-groups can communicate securely without revealing their participation to the wider group.
Anonymous group encryption can be built up from pair-wise encryption for which can be made unconditionally secure via quantum key distribution~\cite{Huang2022}.
Multi-partite entanglement can be used to establish a group key between $N$ users, consuming $(N-1)$ times fewer network resources than equivalent pair-wise entanglement schemes; this task is known as quantum conference key agreement~\cite{Epping2017,Grasselli2019,Murta2020,Proietti2021,Pickston2023}.
More recently, it has been shown that anonymous quantum conference key agreement (AQCKA)~\cite{Hahn2020} obtains a larger efficiency advantage from multi-partite entanglement with a theoretical maximum of $N(N-1)$ for an $N$-user network~\cite{Grasselli2022}.
More so, by increasing the anonymity criteria of this task, an even larger advantage of up to $N(N-1)^2$ can be achieved.

\begin{figure*}[ht!]
    \centering
    \includegraphics{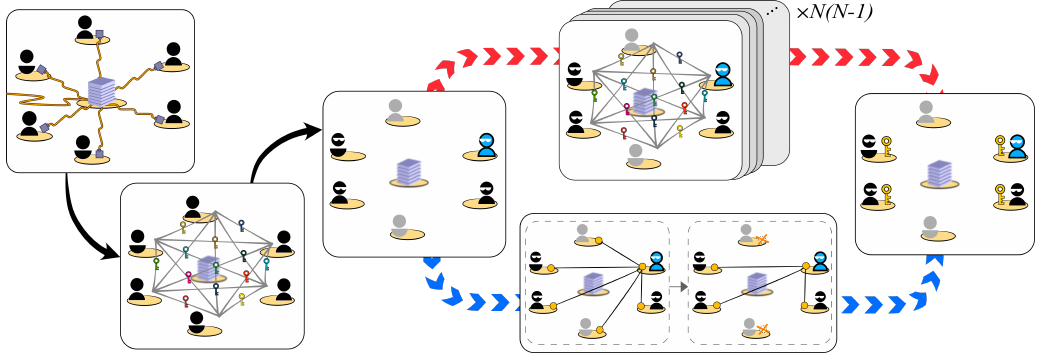}
    \caption{Sketch of the AQCKA protocol. 
    A quantum server distributes bi-partite or multi-partite entangled resources to $N$ users.
    The users perform quantum key distribution with bi-partite entanglement to distill $\binom{N}{2}$ unique pairwise keys.
    The sender (blue) assigns each user their role via anonymous messaging primitives, e.g., keyholders (black) and non-keyholders (gray).
    The anonymous conference key can be established by consuming $N(N-1)$ bi-partite key bits per conference key bit (see red path). 
    Alternatively, by sharing $N$-party GHZ states and applying a pre-shared measurement sequence, the group can directly obtain up to one conference key bit per copy of the GHZ state (see blue path).}
    \label{fig:fullyAQCKAcartoon}
\end{figure*}

There are different ways to realise a quantum network.
A modular approach uses underlying entanglement already present in the network, entangling operations and local operations to dynamically allocate the desired quantum resources to users~\cite{pirker_modular_2018, Hahn2019}. 
A more direct approach to realise a network invokes a central quantum server, which directly generates and distributes entangled resources to required parties~\cite{Huang2022,Wengerowsky2018entanglement,Proietti2021,Avis2023}.
We consider the task of AQCKA with the latter network approach, as illustrated in Fig.\ref{fig:fullyAQCKAcartoon}, where $N$ users are connected to the quantum server.

The task is outlined as follows.
Users form two subsets: $M$ keyholders (including a sender) while remaining $N-M$ as non-keyholders.
The keyholders' aim to establish a conference key, co-ordinated by the sender, while the non-keyholders do not learn the key nor identities of keyholders.
Bi-partite resources are distributed in an all-to-all configuration for pairwise keys to satisfy anonymous subprotocols~\cite{Broadbent2007,Huang2022}.
This ensures anonymity whereby any user adopts the role of sender by first carrying out a subroutine to ensure only one sender is attempting to initialise AQCKA.
Upon success, the sender anonymously notifies only the keyholders of the conference key agreement subroutine, which determines how the shared entanglement is to be used to securely distill the conference key.
Once the roles have been assigned, every round requires users to perform measurements on their resource to either distill a raw key, or check for eavesdropping.
At this stage, users can conduct the remaining steps of the protocol with either bi-partite resources, or multi-partite resources through either a maximally-entangled GHZ state~\cite{Hahn2020,Thalacker2021,Grasselli2022} or a linear cluster~\cite{dejong2022, Ruckle2022}.
The non-keyholders, unaware of who the keyholders are, repeatedly perform disentangling measurements for the multi-partite variant.
Any deviation by non-keyholders, or interaction by an unwanted eavesdropper, will introduce noise that is detected during parameter estimation---a step dedicated to testing the errors in the distributed state, and will lead to the protocol being aborted.
Provided the protocol does not abort, the sender anonymously broadcasts the information the keyholders require to perform error correction on their respective raw keys, followed by a verification round to ensure the keys are identical.
Lastly, the keyholders implement privacy amplification on their key to arrive at the final secret conference key.

In this work, we experimentally perform multi-partite and bi-partite variants of AQCKA by generating a six-qubit photonic GHZ state that is distributed from a central server, forming a six-user network.
To realise each variant, users perform different measurements on the network to distill either Bell pairs for bi-partite resources or GHZ states for multi-partite resources.
The key rate performance, in the asymptotic limit, is evaluated for both variants with two different anonymity conditions in a six-user network.
Using a four-user network and considering finite-key effects, we provide an in-depth analysis of AQCKA based on multi-partite entanglement to highlight the amount of rounds required for performing each subprotocol and the finite-key rate.

\section*{experiment}

\begin{figure}[!b]
    \centering
    \includegraphics[width=\linewidth]{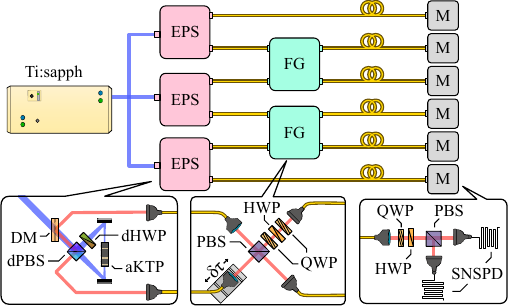}
    \caption{Experimental setup.
    A pulsed Ti:sapph laser pumps three entangled photon-pair sources (EPS).
    Each EPS embeds an aKTP crystal in a Sagnac configuration consisting of a dual-wavelength polarising beamsplitter (dPBS), half-wave plate (dHWP) and dichroic mirror (DM). Diagonally-polarised pump photons propagate clockwise and anti-clockwise to generate orthogonally polarised photon-pairs through type-II spontaneous parametric down-conversion process in the aKTP crystal. The photon pairs are separated by the dPBS, upon which they are coupled into single-mode fibres.
    Two fusion gates (FG) are realised by interfering two photons on a PBS, with a delay stage ($\delta\tau$) to ensure temporal indistinguishability.
    Two quarter-wave plates (QWP) and a HWP compensate for phase effects in the FGs.
    All six photons are then sent to measurement stages (M) featuring a QWP, HWP and PBS, with both outputs coupled to superconducting nanowire single photon detectors (SNSPD).}
    \label{fig:6GHZ}
\end{figure}

We prepare the six-photon GHZ state using three entangled-photon-pair sources as shown in Fig.~\ref{fig:6GHZ}.
The sources exploit type-II parametric down-conversion using aperiodically-poled KTP (aKTP) crystals tailored to produce spectrally separable biphoton states.
This suppresses unwanted spectral correlations created in conventional parametric down-conversion sources allowing us to omit narrowband spectral filtering which improves photon-pair coupling efficiencies~\cite{graffitti_independent_2018,pickston_optimised_2021}.
All three sources are pumped by a mode-locked laser that produces 1.3~ps pulses at a repetition rate of 80~MHz and centre wavelength of 775~nm, creating degenerate telecom photon pairs at 1550~nm.
We create polarisation-entangled photon pairs by optically pumping the aperiodically-poled KTP crystal bidirectionally in a Sagnac configuration \cite{fedrizzi2007wavelength} to produce the $|\Phi^+\rangle$ Bell state with pair production rate of $\sim~2\times10^3$~Hz/mW/s.
Two linear optical fusion gates interfere photons from separate sources in order to create the six-photon GHZ state.
The success of these fusion gates is conditioned on the detection of a single photon in each measurement stage, see Fig.~\ref{fig:6GHZ} for details.
We measure an average six-photon rate of $0.67$~Hz for a pump power of $300$~mW.
We also prepare a four-photon GHZ state, using two sources and one fusion gate, measuring a four-photon rate of $41.6$~Hz at $300$~mW.

To characterise the quality of our photonic GHZ states, we establish a lower-bound fidelity to the ideal state through stabiliser measurements~\cite{Toth2005}.
Our estimated fidelities for the four- and six-photon GHZ states are $0.933(4)$ and $0.825(5)$ respectively at a nominal pump power, see Appendix for further details.
In the AQCKA task, both bi-partite and multi-partite resources are required from the quantum server which is distributing GHZ states in every round of the task.
When a smaller GHZ state or Bell pair is required, non-keyholders locally measure their photon in the Pauli-$X$ basis to disentangle themselves from the network, whilst keyholders measure in the Pauli-$Z$ basis.
For example, to obtain a Bell pair resource in a six-party network, four users measure in $X$ and the remaining two measure in $Z$.
As a result, the generation rate of the Bell pairs is the same as the full-sized GHZ state as post-selecting on six-folds is required for the $N=6$ case.

\section*{results}
\subsection*{Asymptotic-key regime}

We first evaluate the performance of the AQCKA protocol without multi-partite entanglement, denoted AQCKA$_\text{B}$ henceforth and detailed in Appendix, by measuring the noise terms of every Bell pair in the six-user network.
This yields fifteen configurations of user pairs indexed by $q$ and $t$, where $\{q,t\} \in {1,...,6}$ and $t>q$.
In each case we measure the quantum bit error rate ($Q_{Zb}^{q,t}$) and phase error ($Q_{Xb}^{q,t}$) respectively.
Each user pair implements standard BBM92 protocol \cite{Bennett92} to establish bi-partite secret keys, whose asymptotic key rate (AKR) is given by: $\sigma_\text{B}^{q,t} = 1 - h(Q_{Zb}^{q,t}) - h(Q_{Xb}^{q,t})$, where $h(\cdot)$ denotes the binary entropy function.
These key rates dictate the overall performance of AQCKA$_\text{B}$ as the protocol relies exclusively on the bi-partite keys to transmit the conference key directly to the keyholders.
In particular, in the asymptotic limit of infinitely many rounds, the amount of bi-partite keys used for the initial designation of each users' role becomes negligible---which is also the case for AQCKA with multi-partite entanglement.
What impacts the performance of AQCKA$_\text{B}$ is the distribution of the conference key to the intended keyholders: for each conference key bit, the protocol consumes $N(N-1)$ bits from the bi-partite keys, independently of the actual number of keyholders.
The resulting AKR for AQCKA$_\text{B}$ is $r_\text{B}^\infty = 0.0109(2)$ (for more details on the calculation see Appendix).
We remark that the AQCKA approach without multi-partite entanglement does not require anonymity in its quantum phase, i.e., when the bi-partite keys are established, and can thus rely on conventional QKD primitives.
The anonymous distribution of the conference key is then carried out with classical anonymous messaging primitives \cite{Broadbent2007} based on the established bi-partite keys.

Next, we evaluate the performance of AQCKA$_\text{M}$ which uses multi-partite entanglement in the form of six-party GHZ states.
Once users are notified of their roles, non-keyholders measure only in the $X$ basis while keyholders measure in either $Z$ or $X$, following the prescribed test-key sequence, corresponding to key-generation and test rounds respectively.
The group collectively estimate the phase error, $Q_X$, by announcing their $X$ measurement outcomes using anonymous messaging primitives.
Unlike in conventional conference key agreement protocols, the key error rate $Q_Z$, which is defined as the maximum pairwise error, is not estimated during the protocol.
Instead, the keyholders will use a pre-established $Q_Z$ that is representative of the network, which allows fewer anonymous messaging rounds while allowing error correction to take place---see Appendix for a detailed description of AQCKA$_M$ protocol.
For the purposes of providing a complete characterisation of the network, we measure $Q_Z$ for all possible configurations of keyholders.
We obtain the asymptotic key rate of $r^\infty_M = 0.235(12)$, which corresponds to an increase over the AQCKA protocol without multi-parite entanglement of $r^\infty_M/r^\infty_B = 21.6(1.1)$.
Theoretically we expect a ratio of $30$ for our $N=6$ network, however this only holds in the ideal scenario where noise is symmetric.
In practice, channel noise will impose a higher penalty on the AQCKA$_\text{M}$ protocol, as we must operate under the highest link noise to preserve anonymity.

Using the same data set, we evaluated the performance of a related AQCKA task,  denoted fully-AQCKA in~\cite{Grasselli2022}, that demands a stronger anonymity condition where even the keyholders are oblivious to each other (except for the sender who selects the keyholders).
The immediate consequence of this modification is that the keyholders can no longer use a pre-shared conference key to perform efficient multi-party error correction, instead they must use the more costly bi-partite private channels. Similarly, the amount of bi-partite resources consumed by the protocol without multi-partite entanglement increases by an extra factor of $(N-1)$ in order to hide the identity of the sender.
In this scenario, a maximum multi-partite resource advantage of $N(N-1)^2$ can be obtained which for a $N=6$ network is $150$.
However this advantage quickly diminishes with increasing $Q_Z$, due to a larger overhead in the required anonymous messaging for error correction.
Our noise parameters lead to a $r^\infty_\text{fully-M} = 0.0075(6)$ which equates to a multi-partite resource advantage of $3.42(27)$ which is not unexpected owing to the strong $Q_Z$ dependence.
For further information on this scaling and the protocols, see Appendix.

\subsection*{Finite-key regime}

When a finite number of rounds are performed, the correct designation of the users' roles, the conference key security, and the anonymity of the keyholders cannot be attained with certainty. To account for a small probability of failure, $\epsilon$, for each of these three features, we adopt the $\epsilon$-secure framework introduced in~\cite{Grasselli2022} as an extension of the composability framework of QKD~\cite{Portmann_2022}.
In this framework, the conference key length ($\ell$) depends on the total  number of protocol rounds ($L_{tot}$), as well as on the parameter $\epsilon$.
For example, the conference key length of AQCKA$_\text{M}$ with finite-key effects reads:
\begin{equation}\label{eq:fkr}
    \ell = L \cdot (1-p)[1 - h(Q_X + \gamma) - h (Q_Z)] - L \cdot h(p) - C\,, 
\end{equation}
where: $L$ is the number of multi-partite rounds, $p$ is the fraction of the multi-partite rounds that are used for parameter estimation and $\gamma$ is the statistical fluctuation of $Q_X$. We do not consider statistical fluctuations of $Q_Z$ since the accepted error rate is fixed before running the protocol and does not affect the secrecy of the key. Finally, $C>0$ is a constant which includes some of the $\epsilon$ parameters: $C=\log_2(2(N-1)/\epsilon_{EC})+2\log_2(1/2\epsilon_{PA})+N$. Both $\gamma$ and $L$ are functions of $L_{tot}$, $p$, and the $\epsilon$ parameters of the protocol. In parallel, the total security parameter of the protocol depends on the $\epsilon$ parameters and is fixed to $\epsilon_{\rm tot} = 1\times10^{-8}$. We refer the reader to Appendix for more details.

\begin{figure}[t!]
    \centering
    \includegraphics[width=\columnwidth]{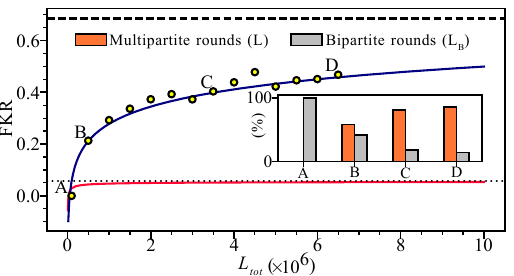}
    \caption{The FKR of the AQCKA$_\text{M}$ protocol, given by $\ell/ L_{tot}$, with $\ell$ given in \eqref{eq:fkr}. The yellow points are experimental data points while the blue line is a simulation of the FKR for fixed noise parameters evaluated from the experiment: $Q_X=0.0304(1)$, $Q_Z=0.01589(5)$ and $Q_{Zb}=0.0144(3)$. The rounds are partitioned into the multi-partite rounds for key generation and testing, and bi-partite rounds for establishing pairwise keys to run classical subprotocols anonymously. The dashed line at the top shows an AKR of the AQCKA$_\text{M}$ protocol of $0.6853(2)$, which is the asymptotic limit of the FKR. The dotted line is the AQCKA$_\text{B}$ AKR of $0.0579(4)$ and the red line is the corresponding FKR given by $\ell'/L_{tot}$, with $\ell'$ reported in Appendix.
    }
    \label{fig:FKRImage}
\end{figure}

We analyse the finite-key rate of AQCKA$_\text{M}$ in a smaller network of $N=4$ users, instead of $N=6$ used for the AKR evaluations, to increase the raw key generation rate with three users being keyholders.
In $L$ multi-partite rounds, the users share four-photon GHZ states to extract shared key bits. In $L_B=L_{tot}-L$ bi-partite rounds, Bell pairs are shared between pairs of users to establish six bi-partite keys used for anonymous classical subprotocols. The finite-key rate (FKR) is defined as the fraction of secure conference key bits established per round ($\ell / L_{tot}$). In Fig.~\ref{fig:FKRImage} we optimized the FKR by numerically maximizing the key length in \eqref{eq:fkr} over the $\epsilon$ parameters and $p$, while keeping the total security parameter fixed.
We measured $Q_X$ and $Q_Z$ to be $0.0304(1)$ and $0.01589(5)$ respectively and use these noise parameters to numerically simulate the FKR, shown as the blue trend line in Fig.~\ref{fig:FKRImage}.
Further, we simulate the AQCKA$_\text{B}$ FKR as a comparison (red line in Fig.~\ref{fig:FKRImage}), whose finite-key length ($\ell'$) is reported in Appendix.
Note that while the AQCKA$_\text{B}$ protocol reaches the expected AKR for symmetric noise across all pairwise channels for low number of rounds, the FKR is on par with AQCKA$_\text{M}$ and then drastically overtaken from $2\times 10^5$ rounds on wards.

In a real scenario, a round constitutes an instance where all users detect one photon.
Therefore, instead of a collecting measurement statistics over some time, we allocate a measurement time sufficient to detect only a single four-photon generation event per round.
The yellow points in Fig.~\ref{fig:FKRImage} show the experimentally measured FKR for increasing $L_{tot}$ up to $6.5$~million rounds.
Each yellow data point has a fixed $Q_Z$, but a different $Q_X$---found by evaluating the average error in the joint-$X$ measurements from a random sample of $Lp$ test rounds.
Note that we account for finite-key effects even in the generation of the pairwise keys with the $L_B$ rounds, where $Q_{Zb}$ is also fixed to the maximum pairwise error rate: $0.0144(3)$.
We highlight four points, labelled A-D, and show in an inset how their respective $L_{tot}$ is decomposed into multi-partite and bi-partite rounds.
At D, with $6.5$~million rounds and an FKR of $0.47$, we obtained a secure and anonymous conference key of approximately $3$~Mbits.

\section*{discussion}

The drastic resource advantage AQCKA$_\text{M}$ offers is observed in networks, and similar to QCKA~\cite{Pickston2023} our main premise is that the multi-node network resources are provided in the background, as might be the case in a future quantum internet.
Despite the considerable resource overhead, with current state-of-the-art photon generation systems it would be faster to generate anonymous conference keys through bi-partite point-to-point connections. 
However, at present such an approach is limited to trusted-node networks.
While the advantage we observed for fully-AQCKA$_\text{M}$ was lower than AQCKA$_\text{M}$, it has a much greater potential as long as the error rates are below $Q_Z<1\%$, which would be the case in an quantum network which incorporates entanglement purification~\cite{pan2001entanglement,Dur2003,Hu2021} at its untrusted nodes.

When analysing AQCKA$_\text{M}$ with finite key effects, we find that the number of rounds required for parameter estimation is saturated by a very small fraction of the total rounds.
This has the benefit of a greater fraction of total rounds being allocated as multi-partite rounds for distilling a raw key, as well as requiring fewer bi-partite rounds for the anonymous classical subprotocols.
Nevertheless, this comes at the cost of a larger statistical correction when estimating $Q_X$, which leads to a larger amount of key reduction in the privacy amplification step. 
Indeed, even in the regime with over 1 million total rounds, the majority of rounds are multi-partite rounds used for key generation, yet a fairly modest FKR is obtained.
This is due to the statistical correction term, $\gamma$, whose value of 0.03 is comparable to the measured $Q_X$ term.

In conventional, non-anonymous, QCKA~\cite{Proietti2021} the statistical correction can be drastically reduced by simply increasing the share of parameter estimation rounds in the multi-partite rounds, i.e. by increasing $p$. 
However, in AQCKA$_\text{M}$, this would also increase the number of bi-partite rounds required to run the anonymous classical subprotocols, which in turn would further decrease the number of multi-partite rounds used for key generation, making such a choice disadvantageous. 
This effect is ascribed to the additional security feature in AQCKA compared to QCKA, i.e. the anonymity of the keyholders.
In the AQCKA$_\text{B}$ protocol, this behaviour is not observed as the pairwise keys are established with conventional QKD schemes, but distilling the conference key anonymously requires $N(N-1)$ more resources than AQCKA$_\text{M}$. 

As quantum networks grow in size, and the number of connecting users increases, there will be demands for anonymity. 
It has already been shown that multi-partite entanglement provides advantages in quantum secret sharing protocols~\cite{Walk2021} and distributed sensing schemes~\cite{liu2021}.
Our work highlights that multi-partite entanglement provides advantages when layers of anonymity are added to group key encryption tasks.
Motivated by this, there may be other quantum-based cryptographic schemes that gain similar resource advantages when anonymity is added---for example in election voting~\cite{PhysRevApplied.18.014005,PhysRevA.94.022333}, distributed parameter estimation~\cite{Shettell2022,Kasai2022,kasai2023anonymous,PhysRevA.105.L010401}, or in multi-party computation~\cite{movahedi2014secure}.
The modifications of such protocols and their specific resource advantage is the focus of future work.

\section*{Acknowledgements}
\noindent This work is supported by the UK Engineering and Physical Sciences Research Council (EP/T001011/1). GM and FG acknowledge funding from the Deutsche Forschungsgemeinschaft (DFG, German Research Foundation): GM is funded under Germany’s Excellence Strategy - Cluster of Excellence Matter and Light for Quantum Computing (ML4Q) EXC 2004/1 - 390534769; FG is funded by the DFG Individual Grant BR2159/6-1.

\newpage
\widetext

\section*{Appendix}
\section{Anonymous conference key agreement protocols}
We outline in Table~\ref{tab:ScenarioABiProtocol}, Table~\ref{tab:ScenarioABiProtocol2}, Table~\ref{tab:ScenarioAMultiProtocol1}, and Table~\ref{tab:ScenarioAMultiProtocol2} the main steps of AQCKA$_\text{B}$, fully-AQCKA$_\text{B}$, AQCKA$_\text{M}$, and fully-AQCKA$_\text{M}$ protocols respectively.
We note that AQCKA$_\text{B}$ and fully-AQCKA$_\text{B}$ are multi-party generalisations of the Anonymous Message Transmission protocol~\cite{Broadbent2007} that includes specific subprotocols and additional error correction procedures. The reader is directed to \cite{Grasselli2022} for further details.

\begin{table}[h!]
\centering
\caption{\textit{AQCKA$_\text{B}$ protocol.} Bell states constitute the protocol's resource and are used to establish pairwise keys amongst all users with a BBM92 protocol~\cite{Bennett92}.
In order to obtain a conference key, the protocol uses $N(N-1)$ bi-partite bits for each bit of conference key.}
\begin{tabular}{l p{11.5cm}}
\hline
\multicolumn{1}{l}{Step} & \multicolumn{1}{l}{Description} \\
\cline{1-2}\\
0 & \emph{Pre-established keys.} Every pair of users in the $N$-user network performs two-party quantum key distribution to obtain pairwise secret keys. In total there are $\binom{N}{2}$ combinations.\\
1 & \textit{Identity designation} (ID). A single Sender is assigned anonymously. The Sender notifies the remaining parties of their roles as keyholder or non-keyholder and informs the keyholders about the identities of the Sender and of the other keyholders. This consumes bi-partite keys to ensure anonymity.\\
2 & The Sender generates a random bitstring of length $L_b$ which is the conference key $\vec{k}_{Conf}$.\\
3 & Each user sends a random bitstring of length $L_b$ (except the Sender) to every other user through the bi-partite channels. The Sender sends $\vec{k}_{Conf}$ to the keyholders and a random bitstring to the non-keyholders. The keyholders identify the string $\vec{k}_{Conf}$ received by the Sender as the conference key.\\
\cline{1-2}
\end{tabular}
\label{tab:ScenarioABiProtocol}
\end{table}

\begin{table}[h!]
\centering
\caption{\textit{Fully-AQCKA$_\text{B}$ protocol.} Bell states are the protocol's resource and are used to establish pairwise keys amongst all users with a BBM92 protocol.
In order to obtain a conference key, the protocol uses $N(N-1)^2$ bi-partite bits for each bit of conference key.}
\begin{tabular}{l p{11.5cm}}
\hline
\multicolumn{1}{l}{Step} & \multicolumn{1}{l}{Description} \\
\cline{1-2}\\
0 & \emph{Pre-established keys.} Every pair of users in the $N$-user network perform two-party quantum key distribution to obtain pairwise secret keys. In total there are $\binom{N}{2}$ combinations.\\
1 & \textit{Identity designation} (ID). A single Sender is assigned anonymously. The Sender notifies the remaining parties of their roles as keyholder or non-keyholder. This consumes bi-partite keys to ensure anonymity.\\
2 & The Sender generates a random bitstring of length $L_b$ which is the conference key $\vec{k}_{Conf}$.\\
3 & The parties repeat the following for $N-1$ times: the Sender anonymously notifies one of the parties and subsequently communicates the conference key if they are a keyholder or a random string if the party is a non-keyholder. In doing so, the Sender remains anonymous. \\
4 & The protocol aborts publicly if any of the keyholders detects a failure in one of the previous steps. This consumes bi-partite keys to ensure anonymity. \\
\cline{1-2}
\end{tabular}
\label{tab:ScenarioABiProtocol2}
\end{table}

\begin{table}[h!]
\centering
\caption{\textit{AQCKA$_\text{M}$ protocol}. Bell pairs and $N$-party GHZ states are used as resources by the protocol.
More precisely, the anonymous subprotocols use Bell pairs while the \textit{quantum measurements} in step $3$ use $N$-party GHZ states.
We note that the anonymous subprotocols, e.g., Parity, Identity Designation, Error Correction, as well as the full protocol itself is established in \cite{Grasselli2022}, where the reader is redirected for more information.}
\begin{tabular}{l p{11.5cm}}
\hline
\multicolumn{1}{l}{Step} & \multicolumn{1}{l}{Description} \\
\cline{1-2}\\
0 & \textit{Pre-established keys}. The protocol assumes that each subset of parties that are potential keyholders have access to a secure pre-shared conference key $\vec{k}_{Pre}$. Moreover, each pair of parties in the network establishes a pairwise key, for a total of $\binom{N}{2}$ pairwise keys, with quantum key distribution.\\
1 & \textit{Identity designation} (ID).
A single Sender is assigned anonymously. The Sender notifies the remaining parties of their roles as keyholder or non-keyholder and informs the keyholders about the identities of the Sender and of the other keyholders. This consumes bi-partite keys to ensure anonymity.\\
2 & \textit{Testing key distribution} (TKD).
The Sender generates a random string of length $L$, where each bit indicates if the round is a test round (bit $1$) or a key-generation round (bit $0$). The probability that a bit is equal to $1$ is given by $p$. The string is compressed into a string $\vec{k}_T$ of $L\cdot h(p)$ bits.
The Sender encodes (with one-time pad) the string $\vec{k}_T$ using $L\cdot h(p)$ bits of the pre-shared key $\vec{k}_{Pre}$ and anonymously broadcast it.
This is done via a public broadcast channel: the keyholders and non-keyholders broadcast random bit-strings of length $L\cdot h(p)$, while the Sender broadcast the encoding of $\vec{k}_T$.
Using $\vec{k}_{Pre}$, the keyholders recover $\vec{k}_T$.\\
3 & \textit{Quantum measurements}. For $L$ rounds, the $N$-party GHZ state is distributed across the multi-partite channels. The Sender and keyholders follow the recipe for measurements indicated by $\vec{k}_T$, where a 1 is a test round and 0 is a key generation round. This translates to either measuring in the X-basis or in the Z-basis, respectively.
The non-keyholders measure in the X-basis in every round.\\
4 & \textit{Testing key broadcast} (TKB). The Sender anonymously broadcasts $\vec{k}_T$ to all users using the (anonymous) Parity protocol~\cite{Broadbent2007} for $L\cdot h(p)$ times, which consumes bi-partite keys to ensure anonymity. For each use of Parity, the Sender inputs a bit of $\vec{k}_T$ while all other users input $0$ such that the users recover a final string $\vec{k}'_T$. \\
5 & \textit{Parameter estimation} (PE). For each test round according to $\vec{k}'_T$, the users performs the Parity protocol as follows: the keyholders and non-keyholders input their measured $X$ outcome. The Sender always inputs a coin flip. The outcome of this is a string $\vec{o}_T$, which the Sender uses to calculate the observed test error rate $Q_X^{obs}$. Verification fails if $Q_X^{obs} + \gamma(Q_X^{obs}) > Q_X + \gamma(Q_X)$, where $\gamma$ is a statistical correction and $Q_X$ is the predefined phase error threshold in the X basis.\\
6 & \textit{Classical post-processing}. According to a predefined pairwise quantum bit error rate $Q_Z$, the Sender anonymously broadcasts  $L\cdot (1-p)\cdot h(Q_Z)+\log\frac{2(N-1)}{\epsilon_{EC}}$ bits of information to all users such that each keyholder can correct their raw key to match the Sender's key. The broadcast is encrypted by consuming bits from the string $\vec{k}_{Pre}$, hence only the keyholders can decrypt it. The protocol aborts if at least one between the error correction and the PE verification failed for any of the keyholders. This information is only available to the keyholders by having each keyholder broadcast a bit encrypted with $\vec{k}_{Pre}$.
Privacy amplification extracts a secure key from the error-corrected keys of the keyholders by applying a two-universal hash function. The resulting conference key has length $l=L\cdot (1-p)\cdot [1-h(Q_X+\gamma(Q_X))]-2\log\frac{1}{2\epsilon_{PA}}$.
\\
7 & \textit{Final secure key}. As $l_c=L\cdot h(p)+L\cdot (1-p)\cdot h(Q_Z)+\log\frac{2(N-1)}{\epsilon_{EC}}+N$ bits of $\vec{k}_{Pre}$ are consumed by the protocol, the net length of the resulting secure conference key is: $\ell= l-l_c$ and the resulting key rate is $\frac{\ell}{L_{tot}}$,where $L_{tot}=L+L_B$ accounts for the rounds used to generate the required pre-shared bi-partite keys.\\
\cline{1-2}
\end{tabular}
\label{tab:ScenarioAMultiProtocol1}
\end{table}

\begin{table}[h!]
\centering
\caption{\textit{Fully-AQCKA$_\text{M}$ protocol}. Bell pairs and $N$-party GHZ states are used as resources by the protocol.
More precisely, the anonymous subprotocols use Bell pairs while the \textit{quantum measurements} in step $3$ use $N$-party GHZ states.
We note that the anonymous subprotocols, e.g., Parity, Identity Designation, Error Correction, as well as the full protocol itself is established in \cite{Grasselli2022}, where the reader is redirected for more information.}
\begin{tabular}{l p{11.5cm}}
\hline
\multicolumn{1}{l}{Step} & \multicolumn{1}{l}{Description} \\
\cline{1-2}\\
0 & \textit{Pre-established keys}. Each pair of parties in the network establishes a pairwise key, for a total of $\binom{N}{2}$ pairwise keys, with quantum key distribution.\\
1 & \textit{Identity designation} (ID).
A single Sender is assigned anonymously. The Sender notifies the remaining parties of their roles as keyholder or non-keyholder.
This consumes bi-partite keys to ensure anonymity. \\
2 & \textit{Testing key distribution} (TKD).
The parties consume bi-partite keys in order for the Sender to anonymously distribute to each other keyholder: a string $\vec{k}_T$, of size $L\cdot h(p)$ bits, encoding the test and key-generation rounds and a string $\vec{r}_l$ that will be used to encode the information about whether the protocol aborts.
\\
3 & \textit{Quantum measurements}. For $L$ rounds, the $N$-party GHZ state is distributed across the multi-partite channels. The Sender and keyholders follow the recipe for measurements indicated by $\vec{k}_T$, where a 1 is a test round and 0 is a key generation round. This translates to either measuring in the X-basis or in the Z-basis, respectively.
The non-keyholders measure in the X-basis in every round.\\
4 & \textit{Testing key broadcast} (TKB). The Sender anonymously broadcasts $\vec{k}_T$ to all users using the (anonymous) Parity protocol~\cite{Broadbent2007} for $L\cdot h(p)$ times, which consumes bi-partite keys to ensure anonymity. For each use of Parity, the Sender inputs a bit of $\vec{k}_T$ while all other users input $0$ such that the users recover a final string $\vec{k}'_T$.\\
5 & \textit{Parameter estimation} (PE). For each test round according to $\vec{k}'_T$, the users performs the Parity protocol as follows: the keyholders and non-keyholders input their measured $X$ outcome. The Sender always inputs a coin flip. The outcome of this is a string $\vec{o}_T$, which the Sender uses to calculate the observed test error rate $Q_X^{obs}$. Verification fails and the protocol aborts if $Q_X^{obs} + \gamma(Q_X^{obs}) > Q_X + \gamma(Q_X)$, where $\gamma$ is a statistical correction and $Q_X$ is the predefined phase error threshold in the X basis.\\
6 & \textit{Classical post-processing}. According to a predefined pairwise quantum bit error rate $Q_Z$, the Sender anonymously broadcasts (with the Parity protocol) $L\cdot (1-p)\cdot h(Q_Z)+\log\frac{2(N-1)}{\epsilon_{EC}}$ bits of information to all users such that each keyholder can correct their raw key to match the Sender's key. If the error correction fails for at least one keyholder, the protocol aborts but this information is encoded with $\vec{r}_l$ and only available to the keyholders.
Privacy amplification extracts a secure key from the error-corrected keys of the keyholders by applying a two-universal hash function. The resulting conference key has length: $\ell=L\cdot (1-p)\cdot [1-h(Q_X+\gamma(Q_X))-h(Q_Z)]-\log\frac{2(N-1)}{\epsilon_{EC}}-2\log\frac{1}{2\epsilon_{PA}}$ is obtained.
\\
7 & \textit{Final secure key}. The final secure conference key rate is given by $\frac{\ell}{L_{tot}}$, where $L_{tot}=L+L_B$ accounts for the rounds used to generate the required pre-shared bi-partite keys.\\
\cline{1-2}
\end{tabular}
\label{tab:ScenarioAMultiProtocol2}
\end{table}

\section{Asymptotic key rates}
We describe here the AKR expressions for the AQCKA task.
The respective AKRs for the protocol with and without multi-partite entanglement are given by:
\begin{align}\label{eq:r}
    \text{r}_M^\infty & =
    1-h(Q_X)-h(Q_Z)\\ \label{eq:rb}
     \text{r}_B^\infty & =
          \frac{1}{\frac{N(N-1)}{\binom{N}{2}} \sum_{q<t}[\sigma^{(q,t)}]^{-1}},
\end{align}
where $Q_X=\frac{1-\left<X^{\otimes N}\right>}{2}$ is the global error rate in the $X$-basis and $Q_Z=\max_{(q,t)}\frac{1-\left<Z_q\otimes Z_t\right>}{2}$ the maximum pairwise quantum bit error rate (QBER) in the Pauli $Z$-basis. Moreover $\sigma^{(q,t)} = 1-h(Q_{Xb}^{(q,t)})-h(Q_{Zb}^{(q,t)})$ denote the rates at which private bits can be established between nodes $q$ and $t$, for $q,t \in \{1\dots N\}$, and $Q_{Xb}^{(q,t)}/Q_{Zb}^{(q,t)}$ are the respective observed quantum bit error rates in the Pauli $X/Z$ basis in a QKD protocol between nodes $q$ and $t$.

To give further insight into each of the expressions above, we first note that Eq.~\ref{eq:r} is the general form for a conference key agreement AKR \cite{Murta2020}.
This is independent of the number of users, only depending on the highest pairwise $Q_Z$ and the global $Q_X$ quantum bit error rate.
In Eq.~\ref{eq:rb} we see a dependence on the number of users which ultimately limits the performance of the bi-partite protocol.
The key growth part of the expression is $(1/N(N-1))\sum_{q<t}[\sigma^{(q,t)}]$, unlike Eq.~\ref{eq:r}, impacted by number of users as well as the channel noise.
This is also irrespective of how many keyholders are designated in a particular run of the protocol.
We note that in \eqref{eq:r} this is distributed symmetrically as we take the worst performing channel across all pairwise users.
The $\binom{N}{2}$ is a normalisation factor to the sum of keys of different pairs of parties $\sum_{q<t}[\sigma^{(q,t)}]^{-1}$.

In our experiment, we perform a complete characterisation of the network by measuring the $Q_Z$ for all possible configurations of keyholders presented in Fig.~\ref{fig:qber}.
Using \eqref{eq:r}, the overall secret key rate is evaluated for the different number of keyholders that can be realised in our six-user network in Table~\ref{tab:AKR}.
A general AQCKA$_\text{M}$ AKR scales independently of the number of keyholders, i.e., the AKR is set according to the lowest performing result in order to preserve anonymity.
Yet due to the difference in $Q_Z$ for each scenario, we see an increase in AKR for larger sets of keyholders.
This increased AKR could be leveraged in scenarios where the total number of keyholders is publicly known, by modifying the protocol accordingly.
\begin{table}[h!]
\centering
\caption{Asymptotic key rates for the AQCKA$_\text{M}$ and AQCKA$_\text{B}$ are indicated by r$_\text{M}^\infty$ and r$_\text{B}^\infty$ respectively.
To evaluate the performance advantage when using multi-partite entanglement, we report the ratios r$_\text{M}^\infty$/r$_\text{B}^\infty$, in all cases r$_\text{B}^\infty$ achieves an AKR of $1.09(2)\%$.
The errors are taken from a Monte-Carlo simulation and reported at one standard deviation.}
\begin{tabular}{l l l l l}
\hline
Keyholders & \multicolumn{1}{l}{Two} & \multicolumn{1}{l}{Three} & \multicolumn{1}{l}{Four} & \multicolumn{1}{l}{Five} \\
\cline{2-5}
r$_\text{M}^\infty$ & 0.235(12) & 0.267(09) & 0.282(08) & 0.296(09)\\
Ratio & 21.6(1.1) & 24.5(9) & 25.3(9) & 27.1(9)\\
\hline
\end{tabular}
\label{tab:AKR}
\end{table}

\section{Fully-AQCKA task}
The fully-AQCKA task has a stronger anonymity requirement, wherein only the sender knows the identities of the keyholders while all other users know only their role.
Thus, even keyholders cannot discover the origin of the message transmission nor the identities of other recipients.
To achieve this, the protocol based on bi-partite entanglement, fully-AQCKA$_\text{B}$, requires a larger resource cost owing to the multi-user anonymous message protocols being repeated $N-1$ more times. Moreover the protocol based on multi-partite entanglement, fully-AQCKA$_\text{M}$, also requires additional use of bi-partite resources that grows with the key length depending on the error rate $Q_Z$.
Below are the AKR expressions for this task,
\begin{align}\label{eq:rf}
     \text{r}_{\rm fully-M}^\infty & =
     \frac{\text{r}_M^\infty}{1 + \frac{N(N-1)}{\binom{N}{2}} h(Q_Z) \sum_{q < t}[\sigma^{(q,t)}]^{-1}}  =
     \frac{\text{r}_M^\infty}{1+h(Q_Z)[\text{r}_{B}^\infty]^{-1}}\\ \label{eq:rfb}
     \text{r}_{\rm fully-B}^\infty & =
     \frac{1}{\frac{N(N-1)^2}{\binom{N}{2}}\sum_{q<t}[\sigma^{(q,t)}]^{-1}}=\frac{\text{r}_B^\infty}{N-1}.
\end{align}
The fully-AQCKA$_\text{M}$ protocol is expected to outperform fully-AQCKA$_\text{B}$, akin to AQCKA task, when employing multi-partite resources, leading to a theoretical resource cost that scales as $N(N-1)^2$, which again neglects loss and accounts for the single Bell state distribution per round.
\begin{table}[h!]
\centering
\caption{Asymptotic key rates for the fully-AQCKA task with fully-AQCKA$_\text{M}$ and fully-AQCKA$_\text{B}$ expressed as r$_{\rm fully-M}^\infty$ and r$_{\rm fully-B}^\infty$ respectively.
To evaluate the performance advantage when using multi-partite entanglement, we report the the ratios r$_{\rm fully-M}^\infty$/r$_{\rm fully-B}^\infty$. The r$_{\rm fully-B}^\infty$ achieved an AKR of $0.218(4)\%$.
To ensure anonymity of each set of keyholders the worst performing pairwise channel is used to set the $Q_Z$
The errors are taken through a Monte-Carlo simulation and reported at one standard deviation.}
\begin{tabular}{l l l l l}
\hline
Keyholders & \multicolumn{1}{l}{Two} & \multicolumn{1}{l}{Three} & \multicolumn{1}{l}{Four} & \multicolumn{1}{l}{Five} \\
\cline{2-5}
$r_{\rm fully-M}^\infty$ & 0.0075(6) & 0.0093(6) & 0.0103(6) & 0.0114(7) \\
Ratio & 3.42(27) & 4.28(26) & 4.74(28) & 5.23(35)\\
\hline
\end{tabular}
\label{tab:fullyAKR}
\end{table}

\begin{figure}[h!]
    \centering
    \includegraphics{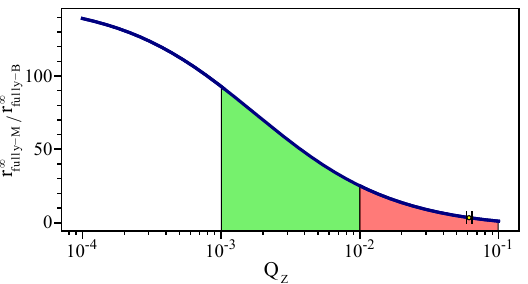}
    \caption{Figure sketching the theoretical curve of the ratio between the fully-AQCKA$_\text{M}$ and fully-AQCKA$_\text{B}$ protocol.
    The yellow point is taken at the worst performing $Q_Z$ for the 6-party network.
    The $Q_{Zb}$ values are taken from Fig.~\ref{fig:qber}(a) where if $Q_Z\leq Q_{Zb}$ then $Q_{Zb}$ is set to $Q_{Zb}:=Q_Z$.
    The green shaded region is the region that is most sensitive to improvements in $Q_Z$.
    The red region is where our $Q_Z$ values lie within.}
    \label{fig:fully-aqcka-curve}
\end{figure}

In Table~\ref{tab:fullyAKR} we present the AKRs of the fully-AQCKA task in all keyholder configurations. 
Similarly to the AQCKA case, the increased key rates for different numbers of keyholders can be leveraged if the total number of keyholders can be publicly known.
Although the fully-AQCKA$_\text{M}$ can achieve an advantage of up to 150 for the $N=6$ case in comparison with a protocol based only on bi-partite resources, our results show that the obtained advantage is lower than that of the AQCKA task.
This is inline with the theoretical result as, in \eqref{eq:rf}, there is a factor of $h(Q_Z)$ that decreases this advantage.
This factor arises as the identities of keyholders are not known hence they cannot use a pre-shared conference key for error correction, instead relying on bi-partite private links and anonymous meassage transmission.
Fig.~\ref{fig:fully-aqcka-curve} shows the theoretical curve obtained for the ration r$_{\rm fully-M}^\infty$/r$_{\rm fully-B}^\infty$ by varying the $Q_Z$ with two regions of interest highlighted.

We find, for a typical range of $0.01\leq Q_Z \leq 0.1$, the multi-partite advantage reaches $\sim25$.
For $Q_Z\leq 0.01$, the scaling becomes non-linear such that improvements on the order of $10^{-3}$ to $Q_Z$ significantly boost the advantage.
It is not atypical that large networks can operate at a $Q_Z$ between $1\%$ to $10\%$ indicated by the red region.
While improvements to $Q_Z$ do increase the multi-partite advantage, the return for leading up to the $1\%$ mark only starts to gain momentum from $\approx 2\%$ onward.
The most sensitive region to $Q_Z$ is below $1\%$, highlighted in green, where slight improvements have the largest gains.
Despite the green region in Fig.~\ref{fig:fully-aqcka-curve} being out of reach for this experiment, it is not unfeasible as mentioned in the main text if, for example, purification protocols~\cite{Bennett1996} or even quantum repeater network architectures~\cite{Briegel1998,lee2022} are used to increase the fidelity of the resource state.

\newpage

\section{Comparison to previous works}

In Table~\ref{tab:comparison2}, we summarise the experiments on anonymous communication with multi-partite quantum resources that have been carried out so far.
We note that, in terms of minimum number of rounds for a non-zero finite key rate, R\"uckle et al.~\cite{Ruckle2022} achieved $L^{\rm min}=1.46 \cdot 10^8$ with a total security parameter of $\epsilon_{\rm tot}=3\cdot 10^{-5}$, while our work achieves $L^{\rm min}=2.87 \cdot 10^5$ with $\epsilon_{\rm tot}=10^{-8}$.
As for the asymptotic key rates for each of the works, Thalacker et al.~\cite{Thalacker2021} performed a fully-AQCKA protocol whose asymptotic key rate tends to zero ($r_{\rm fully-M}^\infty\rightarrow 0$) with a fixed anonymity parameter, while R\"uckle et al.~\cite{Ruckle2022} performed an AQCKA protocol with $r_M^\infty=0.04(6)$. For our work, we performed both the AQCKA and fully-AQCKA protocol whose respective AKR reached $r_M^\infty=0.296(09)$ and $r_{\rm fully-M}^\infty=0.0114(7)$.
\begin{table}[h!]
\caption{For each experiment, we specify (from left to right): the cryptographic task (AQCKA or fully-AQCKA); the type of security proof; the necessary assumptions for the security to hold; whether or not the protocol can tolerate noise; the number $N$ of network users involved.\\ $^*$ By semi-trusted source, we intend that the source of entanglement is still verified by the
protocol and any detected noise is assigned to the eavesdropper. However, the source cannot be
actively malicious and prepare tailored states that would expose the parties’ identities through
their public announcements, at the cost of aborting the protocol.}
\centering
{\small
\begin{tabular}{c c c c c c}
\hline
\parbox[l]{6.5em}{\textbf{Experiments}} & \textbf{Task} & \textbf{Security}& \parbox[c]{13em}{\textbf{Assumptions}} & \parbox[l]{5em}{\textbf{Noise \newline resist.}} & $\mathbf{N}$  \\
\hline
\parbox[c]{6.5em}{\vspace{1ex} Thalacker et al.~\cite{Thalacker2021} } & fully-AQCKA & heuristic~\cite{Unnikrishnan2019,Hahn2020} & \parbox[m]{13em}{\vspace{1ex}$\bullet$ semi-trusted q. source (no active eavesdropping)$^*$ \newline $\bullet$ colluding  dishonest users}  & No & $4$ \\
\hline
\parbox[l]{6.5em}{\vspace{1ex}R\"uckle \newline et al.~\cite{Ruckle2022}} & AQCKA &  $\varepsilon$-security~\cite{dejong2022}& \parbox[m]{13em}{\vspace{1ex}$\bullet$ semi-trusted q. source (no active eavesdropping)$^*$ \newline $\bullet$ non-colluding  dishonest users\vspace{1ex}} & Yes  & $4$ \\
\hline
\parbox[c]{6.5em}{\textbf{This work}} & \parbox[c]{7em}{AQCKA \& \newline fully-AQCKA} & \parbox[c]{8em}{\vspace{1ex} $\varepsilon$-security and \newline $\varepsilon$-anonymity~\cite{Grasselli2022}} & \parbox[m]{13em}{\vspace{1ex} $\bullet$ eavesdropper controls quantum source \newline $\bullet$ eavesdropper can communicate with colluding dishonest users \newline $\bullet$ bounded storage for network users\vspace{1ex}} & Yes & $6$ \\
\hline
\end{tabular}
}
\label{tab:comparison2}
\end{table}

\section{Experimental detail}
The $|\Psi^-\rangle$ states produced by the Bell state sources are locally rotated to obtain $|\Phi^+\rangle$, allowing us to reach a product state such as,
\begin{align*}
|\Psi\rangle = \frac{1}{\sqrt{2}}(|H_1H_2\rangle + |V_1V_2\rangle) \otimes \\ \frac{1}{\sqrt{2}}(|H_3H_4\rangle + |V_3V_4\rangle) \otimes \\ \frac{1}{\sqrt{2}}(|H_5H_6\rangle + |V_5V_6\rangle).
\end{align*}
The aKTP crystals are in the centre of a Sagnac interferometer configuration, which allows erasure of the which-path information.
The quality of each of our sources can be quantified by running a two-qubit tomographic measurement on each of the sources, determining the fidelity and purity.
For sources labelled \{S1, S2, S3\} pumping with 20mW, we report fidelities of \{$98.82\%$, $98.60\%$, $98.60\%$\} and purities of \{$97.98\%$, $97.90\%$, $97.40\%$\}.

The photons are then sent to the circuit consisting of two probabilistic fusion gates~\cite{Browne2005}, each with a probability of success of $1/2$.
As we directly generate Bell states from the sources, our total six-party GHZ baseline probability of success is $1/4$.
With the local rotations on each of the Bell states mentioned prior, as well as the entanglement operations of the fusion gates, we arrive at the six-party GHZ state,
\begin{equation}
    |\Psi^{6}_{GHZ}\rangle = \frac{1}{\sqrt{2}}(|H_1H_2H_3H_4H_5H_6\rangle + |V_1V_2V_3V_4V_5V_6\rangle).
\end{equation}

In order to have a high quality fusion operation, the fusion gate need to be optimised in the spectral, temporal, spatial and polarisation degrees of freedom.
The spectral and polarisation degrees of freedom are bound to the quality of the source, specifically how indistinguishable the photons entering the two ports of the fusion gate are.
The polarisation is optimised though optimising the fidelity with respect to a $|\Psi^-\rangle$ state.
The spectral degree of freedom is optimised by performing a two-photon interference measurement on the fusion gate, with the two inputs being the two outputs of a source, and temperature tuning the aKTP crystal.
The remaining degrees of freedom, temporal and spatial, are determined by the fusion gate itself.
Temporal optimisation is performed by spatially moving one of the inputs back and fourth, scanning the range in which the difference in delay between both photons entering the PBS is minimum.
Spatial optimisation is done by precise alignment of the input and output couplers with respect to the PBS, such that the two paths the photons can go through are maximally overlapped.

\begin{figure*}[h!]
    \centering
    \includegraphics[width=\textwidth]{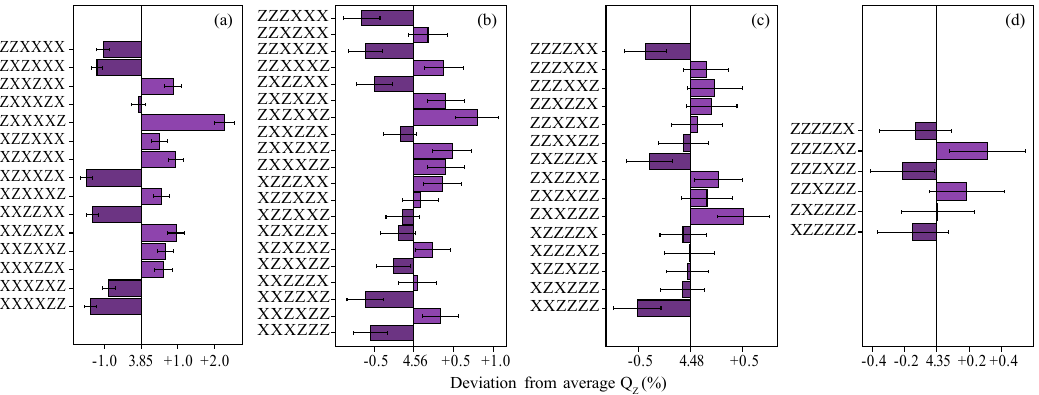}
    \caption{The measured $Q_Z$ for all permutations in the (a) two-user, (b) three-user, (c) four-user and (d) five-user network scenarios in the 300mW pump power regime.
    The labels are read from left to right for each qubit, for example users 1 and 2 measuring in $Z$ while users 3, 4, 5 and 6 measure in $X$ corresponds to $Z_1,Z_2,X_3,X_4,X_5,X_6 = ZZXXXX$.
    Error bars are estimated via Monte-Carlo simulation assuming Poissonian statistics and reported at one standard deviation.}
    \label{fig:qber}
\end{figure*}

The pump power going into each of the sources varies the overall performance in terms of key rates.
While increasing the pump power increases the six-fold rate, it has a negative effect on the quality of the GHZ state due to multi-photon events.
A mask of multi-photon events greater than six is taken from the FPGA, which records all possible detector patterns, and subtracted from the six-fold pattern we post select on.
This allows the removal of a subset of multi-photon events from the data.
The $Q_Z$ in the Z basis, for $300$mW of pump power on each source, for each network scenario are presented in Fig.~\ref{fig:qber}.
Note that users can remove themselves from the network by measuring in $X$.

\newpage

\section{State quality}
In order to determine a lower bounded fidelity of the six-party GHZ and four-party GHZ presented as the resource state in this work, an entanglement witness can be used on two measurement settings \cite{Toth2005}.
Through using the joint-X and joint-Z expectation values, we can consider the following stabilisers:
\begin{align}
    S_1^{(GHZ_N)} := & \prod^N_{k=1}X_{k}\\
    S_k^{(GHZ_N)} := & Z_{(k-1)}Z_k; \ k\in \{2,3,\dots,N\},
\end{align}
where Pauli $X_k$ and $Z_k$ operate on the $k^{th}$ qubit and $S_k^{(GHZ_N))}|GHZ_N\rangle = |GHZ_N\rangle$ holds. This can be used in the genuine $N$-qubit entanglement witness, a witness close to the GHZ states, through
\begin{equation}
    \mathcal{W}=3\cdot \mathbb{I} - 2\Bigg[\frac{S_1^{(GHZ_N)}+\mathbb{I}}{2} + \prod^N_{k=2}\frac{S_k^{(GHZ_N)}+\mathbb{I}}{2}\Bigg].
\end{equation}
From this expression, the lower bounded fidelity can be extracted through
\begin{equation}
    F=\frac{1-\mathcal{W}}{2}.
\end{equation}

We find that for a pump power of 200mW, we achieve a lower bounded fidelity for the six-party GHZ state of $F=0.823(5)$ and for 300mW $F=0.784(4)$.
The noise terms can be seen straight from the expectation values of the observables.
For those that go into the witness expression: $<X>^{\otimes 6}=(0.858(2), 0.823(1))$ and $<Z>^{\otimes 6}=(0.941(3), 0.933(3))$ for 200mW and 300mW respectively.

For the four-party GHZ, we achieve a lower bounded fidelity of $F=0.933(4)$ for a pump power of 40~mW.
The noise terms for this state are $<X>^{\otimes 4} =  0.946(2)$ and $<Z>^{\otimes 4} = 0.974(2)$.

\section{Finite-key analysis}
The finite-key analysis of the AQCKA$_\text{M}$ protocol is conducted on a smaller network of four users, sharing a 4-party GHZ state. The conference key length of the AQCKA$_\text{M}$ protocol is given by:
\begin{equation}\label{eq:fkrsupp}
    \ell = L(1-p)[1-h(Q_X + \gamma (Q_X)) - h (Q_Z)] - \log_2\frac{2(N-1)}{\epsilon_{EC}}-2\log_2\frac{1}{2\epsilon_{PA}}-Lh(p)-N,
\end{equation}
where $\gamma$ is a statistical correction that depends, among other things, on the number of rounds $L$ where a GHZ state is distributed and on the magnitude of the error rate that is being corrected.
The AQCKA$_\text{M}$ protocol requires pairwise channels for the anonymous subprotocols, which also have their distinct finite-key effects and hence finite-key rate. As such, a pairwise channel's key length is,
\begin{equation} \label{eq:ellB}
    \ell_B = \left(\frac{L_{tot}-L}{\binom{N}{2}}\right)(1-p')[1-h(Q_{Xb} + \gamma(Q_{Xb})) - h(Q_{Zb})]-\log_2\frac{2}{\epsilon'_{EC}}-2\log_2\frac{1}{2\epsilon'_{PA}}.
\end{equation}
According to the composable security paradigm, the total $\epsilon$ security parameter of AQCKA$_\text{M}$, due to the finite-key effects affecting the bi-partite keys and the conference key, is given by:
\begin{equation}
    \epsilon_{tot} = 2^{-r_V} + (N - 1) \epsilon_{ enc} + \epsilon_{ EC} + 2\epsilon_{x} + \epsilon_{ PA} + \frac{N(N-1)}{2}(\epsilon'_{ EC} + 2\epsilon'_{ x} + \epsilon'_{ PA}), \label{eq:totalsec}
\end{equation}
where the non-primed parameters refer to the security of the conference key due to the multi-partite rounds. In particular, $r_V$ is the number of rounds dedicated to running the Veto subprotocol \cite{Broadbent2007}, a  primitive present in all the AQCKA protocols analyzed in this paper;  while $\epsilon_{ EC}$, $\epsilon_{ x}$, and $\epsilon_{ PA}$ are related to the failure probability of error correction, parameter estimation, and privacy amplification, respectively, of the conference key. Finally, $\epsilon_{ enc}$ is related to the failure probability of a classical encoding scheme used in the multi-partite part of AQCKA$_\text{M}$. The primed parameters in \eqref{eq:totalsec} are relative to the security of the $N(N-1)/2$ bi-partite QKD protocols that are used to establish all the bi-partite keys.

A set of initial measurements are taken to evaluate the error rates in the different configurations.
Table~\ref{tab:finite_data} shows a tally of the number of rounds experimentally obtained for the finite key analysis and the associated errors, excluding the initiation rounds for channel estimates.
\begin{table}[h!]
    \centering
    \begin{tabular}{l l l }\hline
 \textbf{Configuration}& \textbf{Number of rounds} & \textbf{Error}\\\hline
         ZXZZ&  5624092&  0.01598(5)\\ \hline 
         ZZXX&  154714&  0.0100(2)\\ \hline 
         ZXXZ&  154057&  0.0144(3)\\ \hline
 XXZZ& 152669& 0.0120(2)\\\hline
 ZXZX& 151794& 0.0120(3)\\\hline
 XZXZ& 154872& 0.0122(3)\\\hline
 XZZX& 134390& 0.0132(3)\\\hline
 XXXX& 15575& 0.0304(1)\\\hline
    \end{tabular}
    \caption{Table of experimentally obtained results for the finite key expression Eq.~\ref{eq:fkrsupp}. The measurement time for ZXZZ is around 115 hours, the two user configurations being around 3 hours each and XXXX taking around 19 minutes.}
    \label{tab:finite_data}
\end{table}
In an actual run of the protocol, precursory knowledge of the channel error rates would be known.
For this, initiation rounds were run for each of the configurations.
To highlight the point of this, for the 3 user scenarios the measured highest error rates for $\{XZZZ, ZXZZ, ZZXZ, ZZZX\}$ were $\{0.016(1), 0.013(1), 0.016(2), 0.014(2)\}$ respectively and were each measured for around 5 minutes.
The highest error rate came from the $ZZXZ$ scenario, so the error correction code would be set at this rate --- note that this is not the scenario used in the finite-key analysis, the configuration used for our finite-size AQCKA implementation was $ZXZZ$, and yet the users in all other possible scenarios remain anonymous.

In Fig.~3 of the main text, we optimize the conference key length in \eqref{eq:fkrsupp} over all the security parameters in \eqref{eq:totalsec}, for a fixed total number of rounds $L_{tot}$ and fixed $\epsilon_{tot}$. In particular, the optimal allocation of the total set of rounds ($L_{tot}$) into multi-partite rounds ($L$) and bi-partite rounds ($L_B=L_{tot}-L$) is obtained by requiring that the secret key length $\ell_B$ of each pairwise key is just long enough to cover the needs of each pairwise private channel used by the classical anonymous subprotocols. Now, the total number of pairwise key bits required by the identity designation (ID), testing key broadcast (TKB) and parameter estimation (PE) subprotocols are, respectively:
\begin{align}
    ID = N^2(N-1)(3r_V+|F(\vec{d_t})|)\label{IDbits}\\
    TKB = L\cdot h(p)\cdot N(N-1)\label{TKBbits}\\
    PE = L\cdot p\cdot N(N-1)\label{PEbits},
\end{align}
where $|F(\vec{d_t})|=N-1+\log_2N+2[\log_2(N-1+\log_2N)+\log_2(1/\epsilon_{enc})]$ is an encoded string used to ensure the success of the $ID$ subprotocol, see \cite{Grasselli2022} for further details. Hence, we find the optimal $L$ by solving the following equation for $L$:
\begin{equation}
    \ell_B = \frac{ID+TKB+PE}{\binom{N}{2}}. \label{eqL}
\end{equation}
Note that the function $\gamma$ in \eqref{eq:ellB} is defined implicitly \cite{Grasselli2022}, hence the solution for $L$ is numerical.

Once the optimal value for $L$ is obtained by solving \eqref{eqL}, the total number of bi-partite rounds $L_B$ is obtained as: $L_B=L_{tot}-L$. This ensures that the parties have just enough bi-partite rounds to establish pairwise keys whose length is sufficient to run the anonymous subprotocols. At the end of AQCKA$_\text{M}$, the pairwise keys are completely depleted.

Furthermore, it is interesting to analyze the different demands for bi-partite key bits of the three anonymous subprotocols, namely ID, TKB, and PE, and how this affects the total number of bi-partite rounds $L_B$. For example, the ID subprotocol requires a number of bits that is independent of $L$, while the TKB and PE subprotocols require a number of bits linear in $L$. To analyze how such demands affect the allocation of rounds to bi-partite rounds, we can partition $L_B$ in proportion to the number of bi-partite bits required by each subprotocol:
\begin{align}
    L_B= L_{ID} + L_{TKB} + L_{PE},
\end{align}
where we defined:
\begin{align}
    L_{ID} &= \frac{ID}{ID+TKB + PE} L_B \\
    L_{TKB} &= \frac{TKB}{ID+ TKB + PE} L_B \\
    L_{PE} &= \frac{PE}{ID+TKB + PE} L_B.
\end{align}

\begin{figure}[ht!]
    \centering
    \includegraphics[width=0.5\columnwidth]{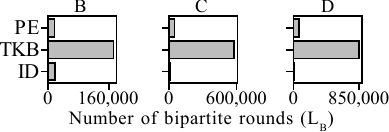}
    \caption{Figure of bi-partite rounds used for increasing number of rounds in the AQCKA$_\text{M}$ protocol. The titles B-D highlight the same points indicated in Fig.~3 of the main text. The A title is not shown as there is no secure key rate for the number of total rounds. The bi-partite rounds are partitioned into Testing Key Broadcast (TKB), which is a subprotocol that anonymously broadcasts the testing key to all users, Parameter Estimation (PE) which computes the noise parameters from the data of the test rounds, and Identity Designation (ID) which establishes the identity of sender and keyholders.}
    \label{fig:NumbOfRounds}
\end{figure}
In Fig.~\ref{fig:NumbOfRounds} we highlight the experimental points B, C and D to analyse how multi-partite and bi-partite rounds are optimally allocated for increasing values of $L_{tot}$.
For points B to D, as more rounds become available, the optimal allocation clearly favours multi-partite rounds over bi-partite rounds. In particular, the bi-partite subprotocols PE and TKB require a number of pairwise key bits that is proportional to $L\cdot p$ and $L\cdot h(p)$, respectively. The fraction of test rounds becomes negligible in the asymptotic limit ($p \to 0$), the ratio $L/L_\text{B}$ increases from B to D, until the finite key rate attains the asymptotic key rate for an infinite number of rounds. For a full breakdown of what constitutes the different yellow datapoints in Fig.~3 in the main text, see Table~\ref{tab:finite_data_2}.
\begin{table}[h!]
    \centering
    \begin{tabular}{l l l l l l}\hline
    $\mathbf{L_{tot}}$ & \textbf{L} & $\mathbf{L_B}$ & \textbf{Key rate} & \parbox[m]{4em}{\vspace{1ex}\textbf{Pairwise key rate}} & \textbf{Qx} \\\hline
         100,000 (A)    & 0      & 100,000 & 0      & 0          & 0.5     \\ \hline
         500,000 (B) & 292548 & 207452 & 0.213054  & 0.502661 & 0.022288 \\ \hline 
         1,000,00 & 688239 & 311761 & 0.292418 & 0.522411 & 0.025813 \\ \hline 
         1,500,000 & 1102995 & 397005 & 0.336624 & 0.537964 & 0.026354 \\ \hline
         2,000,000 & 1537304 & 462696 & 0.373163 & 0.558237 & 0.026185 \\\hline
         2,500,000 & 1972553 & 527447 & 0.392900 & 0.561018 & 0.026558 \\\hline
         3,000,000 & 2415994 & 584006 & 0.372665 & 0.559511 & 0.035131 \\\hline
         3,500,000 (C) & 2854096 & 645904 & 0.403610 & 0.557691 & 0.031167 \\\hline
         4,000,000 & 3307700 & 692300 & 0.438673  & 0.568614 & 0.025790 \\\hline
         4,500,000 & 3769944 & 730056 & 0.478092 & 0.577090 & 0.020521 \\\hline
         5,000,000 & 4206860 & 793140 & 0.422365 & 0.578849 & 0.034222 \\\hline
         5,500,000 & 4677707 & 822293 & 0.446553 & 0.576451 & 0.030473 \\\hline
         6,000,000 & 5105448 & 894552 & 0.451223 & 0.577093 & 0.030681 \\\hline
         6,500,000 (D) & 5566935 & 933065 & 0.468048 & 0.582763 & 0.028282 \\\hline
    \end{tabular}
    \caption{Table of the data used to form the points in Fig.~3 of the main text. The $Q_X$ are taken from the joint-$X$ measurement dataset at random of lengths of $L\cdot h(p)$. The quantum bit error rates are constant and hence not shown in the table.}
    \label{tab:finite_data_2}
\end{table}

For comparison, in Fig.~3 of the main text we also report the FKR of the AQCKA$_\text{B}$ protocol ($\ell'/L_{tot}$), which exclusively relies on bi-partite resources. In this case, the conference key length $\ell'$ is obtained by equating the total number of pairwise key bits required by AQCKA$_\text{B}$:
\begin{align}
    ID + N(N-1) \cdot \ell'
\end{align}
to the total number of pairwise secret key bits established by every party pair:
\begin{align}
    \binom{N}{2} \left[\left(\frac{L_{tot}}{\binom{N}{2}}\right)(1-p')[1-h(Q_{Xb} + \gamma(Q_{Xb})) - h(Q_{Zb})]-\log_2\frac{2}{\epsilon'_{EC}}-2\log_2\frac{1}{2\epsilon'_{PA}}\right].
\end{align}
By solving for $\ell'$, we obtain the finite conference key length of the AQCKA$_\text{B}$ protocol:
\begin{align}
    \ell'= \frac{1}{2} \left[\left(\frac{L_{tot}}{\binom{N}{2}}\right)(1-p')[1-h(Q_{Xb} + \gamma(Q_{Xb})) - h(Q_{Zb})]-\log_2\frac{2}{\epsilon'_{EC}}-2\log_2\frac{1}{2\epsilon'_{PA}}\right] - \frac{ID}{N(N-1)}
\end{align}
where $ID$ is given in \eqref{IDbits}.

\end{document}